\documentclass[aps,preprint]{revtex4}%
\usepackage{amsfonts}
\usepackage{amsmath}
\usepackage{amssymb}
\usepackage{subfigure}
\usepackage{graphicx}
\usepackage{color}%

\begin{document}
\title{Chaos bound and its violation in the torus-like black hole}
\author{}
\author{Rui Yin$^{a,b}$}
\email{yrphysics@126.com}
\author{Jing Liang$^{a,b}$}
\email{ljphysics@163.com}
\author{Benrong Mu$^{a,b}$}
\email{benrongmu@cdutcm.edu.cn}
\affiliation{$^{a}$ Center for Joint Quantum Studies, College of Medical Technology,
Chengdu University of Traditional Chinese Medicine, Chengdu, 611137, PR China}
\affiliation{$^{b}$ Center for Theoretical Physics, College of Physics, Sichuan University,
Chengdu, 610064, PR China}
\begin{abstract}
In this paper, we have studied the variation of the chaos bound in two regions of the torus-like black hole, i.e., the region close to the black hole horizon and the region at a certain distance from the black hole horizon. The angular momentum of the particle affects the effective potential and influences the magnitude of the chaotic behavior of the particle. Therefore, the angular momentum of particle is important in the study. The angular momentum of a particle not only affects the particle equilibrium orbital position, but also affects the Lyapunov exponent. As the angular momentum of the particle increases, the particle equilibrium position gradually moves away from the black hole horizon. In the near black hole horizon region, the chaos bound is not violated, however, at the far black hole horizon region, the chaos bound is violated. In addition unlike the charged AdS black hole which has a spherical topology of the horizon, the torus-like black hole has a toroidal topology of the horizon.

\end{abstract}

\maketitle
\tableofcontents{}

\bigskip{}



\section{Introduction}
\label{sec:A}
Chaos is a seemingly random, chance and irregular motion that occurs only in non-linear and non-accumulable dynamical systems, which are sensitive to initial conditions \cite{intro-Sprott:2003jkr,intro-Ott:2002byv,intro-Brown:1996dyw}. Since the tiny errors in chaotic motion grow rapidly with time, the motion at this point can be quite different from what it would be without these errors. This means that long-term prediction of chaotic motion in general is very difficult. It also means that chaotic motion has many new properties that the usual dynamical systems do not have. This has led to widespread interest in the study of chaos in various areas of physics.

In general relativity, the geodesic motion of particles in a generic Kerr-Newman black hole spacetime \cite{intro-Carter:1968rr} is integrable and there is no chaos in this system. In order to study chaotic motion in general relativity and to ensure that the dynamical system describing the motion of the mass is integrable, it is necessary to resort to some spacetime with a complex geometry or to introduce some additional interactions. Along this spirit, the chaotic motions of particles have been studied in the multi-black hole spacetimes \cite{intro-Dettmann:1994dj,intro-Hanan:2006uf}, the perturbed Schwarzschild spacetime \cite{intro-Bombelli:1991eg,intro-Sota:1995ms,intro-Witzany:2015yqa}, and in the non-standard Kerr black hole spacetime described by MankoNovikov metric \cite{intro-Contopoulos:2011dz,intro-Dubeibe:2007hba,intro-Gueron:2002jt}. The study of chaotic behavior of the geodesic motion of particles has now involved several spacetime contexts, and the main interest of researchers in these systems is to use and further develop coordinate invariant descriptions and metrics of chaotic behavior to make them applicable to general relativity where space and time are not absolute. It has been recently shown that using the Melnikov method to identify chaotic behavior in geodesic motion perturbed by the minimum length effect around a Schwarzschild black hole, there is Smale horseshoes chaotic structure in the phase space \cite{Guo:2020xnf}. Based on the Melnikov method, the existence of a critical amplitude affecting temporal chaos is demonstrated by studying the thermodynamic chaos of RN-AdS black holes immersed in Perfect Fluid Dark Matter, while spatial chaos is always present regardless of the perturbation intensity \cite{Zhou:2022eft}. It has also been tentatively proved that the chaotic behavior of particles near the black hole has quantum gravitational effects \cite{Guo:2020pgq}. In addition, chaotic phenomenon was also investigated for the pinning particles in Kerr spacetime \cite{intro-Han:2008zzf}. More interestingly, chaos in loop string dynamics has been found in the asymptotically flat Schwarzschild black hole spacetime \cite{intro-Frolov:1999pj}, in the AdS-Schwarzschild black hole \cite{intro-Zayas:2010xwj} and in the AdS-Gauss-Bonnet black hole spacetime \cite{intro-Ma:2014aha} after the introduction of loop strings instead of point particles.

In recent studies, a number of violations of the chaos bound have been discovered \cite{intro-Zhao:2018wkl,intro-Kan:2021blg,intro-Gwak:2022xje,intro-Lei:2020clg,intro-Lei:2021koj}. The static equilibrium of charged probe particles around a black hole can be provided by the Lorentz force. In Ref. \cite{intro-Zhao:2018wkl}, Zhao et al. considered the contribution of the sub-leading terms in the expansion of the near-horizon regions and investigated the chaotic bound in the near-horizon regions using the effective potential. It is found that the Reissner-Nordstrom and Reissner-Nordstrom-AdS black holes satisfy this bound, which is violated in a large number of charged black holes. In their study, they only considered the radial contribution. In fact, since angular momentum affects the effective potential and increases the magnitude of the chaotic behaviour of the particles, angular momentum also has an effect on the Lyapunov exponent. In consideration of the above, Lei et al. again studied the chaos bound in the near-visible region of Reissner-Nordstrom and Reissner-Nordstrom-AdS black holes \cite{intro-Lei:2021koj}. It is found that the bound is violated in the near-visible region when the angular momentum of the charge and particles of the black hole is large. In rotating charged black holes, the existence of a violation of the bound was also found by calculations of the effective potential \cite{intro-Kan:2021blg,intro-Gwak:2022xje}.

In this paper, we study the effect of the angular momenta of charged particles on the chaos bounded by the circular motion of particles around a torus-like black hole. The concept of a torus black hole was first introduced in the literature \cite{intro-Huang:1995zb}. Unlike other black holes, the topology of this black hole spacetime is $S\times S\times M^{2}$. This has inspired many studies of torus-like black holes \cite{intro-Zhao:2004qf,intro-Hong:2020zcf,intro-Liang:2021elg,intro-Han:2019kjr,intro-Li:2006kh,intro-Feng:2021vey,intro-Gao:2003ch,intro-Sharifian:2017mmf}.

The rest of this paper is organized as follows. In Sec. \ref{sec:B}, the exact calculation by Jacobi matrix yields the value of the Lyapunov exponent. In Sec. \ref{sec:C}, the variation of the chaos bound of the torus-like black hole is analyzed by numerical calculations, focusing on the regions close to and at a certain distance from the black hole horizon. In Sec. \ref{sec:D}, we devote our discussion to our conclusions.

\section{Lyapunov exponent in the torus-like black hole}
\label{sec:B}
In this subsection, taking into account the motion of charged particles on the equatorial plane of the torus-like black hole, we focus on the calculation of the eigenvalues of the Jacobi matrix, which leads to a general expression for the Lyapunov exponent. The metric of the black hole is
\begin{equation}
ds^{2}=-F(r)dt^{2}+N^{-1}(r)dr^{2}+C(r)d\theta^{2}+D(r)d\phi^{2},
\label{eqn:P21}
\end{equation}
 where the electromagnetic potential $A_{\mu}=A_{t}dt$. When a charged particle moves around a black hole, its Lagrangian can be expressed as
 \begin{equation}
\mathcal{L=\frac{\mathrm{1}}{\mathrm{2}}}(-F\dot{t}^{2}+\frac{\dot{r}^{2}}{N}+D\dot{\phi}^{2})-qA_{t}\dot{t},
\label{eqn:P22}
\end{equation}
here $\dot{x}^{\mu}=\frac{dx^{\mu}}{d\tau}$, where $\tau$ is the proper time. With the aid of the relevant definition of generalized momentum ($\pi_{\mu}=\frac{\alpha\mathcal{L}}{\alpha\dot{x}}$), thus the energy ($E$) and angular momentum ($L$) of the particle can be obtained
\begin{equation}
\begin{aligned}
&E=-\pi_{t}=F\dot{t}+qA_{t},\\
&L=\pi_{\phi}=D\dot{\phi},\\
\end{aligned}
\label{eqn:P23}
\end{equation}
and
\begin{equation}
\pi_{r}=\frac{\dot{r}}{N}.
\label{eqn:P24}
\end{equation}

Then the Hamiltonian quantity of the particle is
\begin{equation}
H=\frac{-(\pi_{t}+qA_{t})^{2}+\pi_{r}^{2}FN+\pi_{\phi}^{2}D^{-1}F}{2F},
\label{eqn:P25}
\end{equation}
using the Hamiltonian quantities of the particles, the radial coordinates and radial momentum versus time can be obtained separately
\begin{equation}
\frac{dr}{dt}=\frac{\dot{r}}{\dot{t}}=-\frac{\pi_{r}FN}{\pi_{t}+qA_{t}},
\label{eqn:P26}
\end{equation}
\begin{equation}
\frac{d\pi_{r}}{dt}=\frac{\dot{\pi_{r}}}{\dot{t}}=-qA_{t}^{\prime}+\frac{1}{2}[\frac{\pi_{r}^{2}FN^{\prime}}{\pi_{t}+qA_{t}}+\frac{(\pi_{t}+qA_{t})F^{\prime}}{F}-\frac{\pi_{\phi}^{2}D^{-2}D^{\prime}F}{\pi_{t}+qA_{t}}],
\label{eqn:P27}
\end{equation}
where
\begin{equation}
\begin{aligned}
&\dot{t}=\frac{\partial H}{\partial\pi_{t}}=-\frac{-(\pi_{t}+qA_{t})}{F},\\
&\dot{\pi_{t}}=-\frac{\partial H}{\partial t}=0,\\
&\dot{r}=\frac{\partial H}{\partial\pi_{r}}=\pi_{r}N,\\
&\dot{\phi}=\frac{\partial H}{\partial\pi_{\phi}}=\frac{\pi_{\phi}}{D},\\
&\dot{\pi_{\phi}}=-\frac{\partial H}{\partial\phi}=0,\\
&\dot{\pi_{r}}=-\frac{\partial H}{\partial r}=-\frac{1}{2}[\pi_{r}^{2}N^{\prime}-\frac{2qA_{t}^{\prime}(\pi_{t}+qA_{t})}{F}+\frac{(\pi_{t}+qA_{t})^{2}F^{\prime}}{F^{2}}-\pi_{\phi}^{2}D^{-2}D^{\prime}],\\
\end{aligned}
\label{eqn:P28}
\end{equation}
hereby, $"\prime"$ denotes the derivative with respect to $t$. The equation $g_{\mu\nu}\dot{x}^{\mu}\dot{x}^{\nu}=\eta$ defines the normalization of the four velocities of a particle, when $\eta=0$, corresponds to the case of a photon, and when $\eta=-1$, represents the case of a massive particle. In this paper, the particle is charged. Next, the constraints can be obtained from the normalization and metric as follows
\begin{equation}
\pi_{t}+qA_{t}=-\sqrt{F(1+\pi_{r}^{2}N+\pi_{\phi}^{2}D^{-1})},
\label{eqn:P29}
\end{equation}
using this constraint, we can obtain
\begin{equation}
\frac{dr}{dt}=\frac{\pi_{r}FN}{\sqrt{F(1+\pi_{r}^{2}N+\pi_{\phi}^{2}D^{-1})}},
\label{eqn:P210}
\end{equation}
\begin{equation}
\frac{d\pi_{r}}{dt}=-qA_{t}^{\prime}-\frac{\pi_{r}^{2}(FN)^{\prime}+F^{\prime}}{2\sqrt{F(1+\pi_{r}^{2}N+\pi_{\phi}^{2}D^{-1})}}-\frac{\pi_{\phi}^{2}(D^{-1}F){}^{\prime}}{2\sqrt{F(1+\pi_{r}^{2}N+\pi_{\phi}^{2}D^{-1})}}.
\label{eqn:P211}
\end{equation}

Next, consider the Lyapunov exponent, in whose acquisition the effective potential of the particle plays an important role \cite{intro-Zhao:2018wkl, intro-Kan:2021blg, intro-Gwak:2022xje}.

The Liapunov exponent can be obtained from the eigenvalue of a Jacobi matrix in the phase space. Considering the motion of the particle in the equilibrium orbit, a condition, namely $\pi_{r}=\frac{d\pi_{r}}{dt}=0$, is needed to constrain the trajectory of the particle. According to literature \cite{Gao:2022ybw}, using the eigenvalues and constraints mentioned above, the specific expression for the Lyapunov exponent can be obtained as follows
\begin{equation}
\lambda^{2}=\frac{1}{4}\frac{N(F^{\prime}+\pi_{\phi}^{2}(D^{-1}F)^{\prime})^{2}}{F(1+\pi_{\phi}^{2}D^{-1})^{2}}-\frac{1}{2}N\frac{F^{\prime\prime}+\pi_{\phi}^{2}(D^{-1}F)^{\prime\prime}}{1+\pi_{\phi}^{2}D^{-1}}-\frac{qA_{t}^{\prime\prime}FN}{\sqrt{F(1+\pi_{\phi}^{2}D^{-1})}}.
\label{eqn:P212}
\end{equation}

From the above equation, it can be observed that the contribution of angular momentum to the Lyapunov exponent is not negligible. The surface gravity is $k=\frac{F^{\prime}(r)}{2}$.

\section{Chaos bound and its violation in the torus-like black hole}
\label{sec:C}

\begin{table}[htb]
\begin{tabular}{|l|lllllll|}
\hline
       & \multicolumn{7}{l|}{$L$ }                                                                                                                                                                            \\ \hline
       & \multicolumn{1}{l|}{0}       & \multicolumn{1}{l|}{1}       & \multicolumn{1}{l|}{2}       & \multicolumn{1}{l|}{3}       & \multicolumn{1}{l|}{5}       & \multicolumn{1}{l|}{10}      & 20      \\ \hline
$Q=0.15$ & \multicolumn{1}{l|}{0.408198}  & \multicolumn{1}{l|}{0.413145} & \multicolumn{1}{l|}{0.426882} & \multicolumn{1}{l|}{0.446916} & \multicolumn{1}{l|}{0.496594}  & \multicolumn{1}{l|}{0.628721} & 0.855718 \\ \hline
$Q=0.20$ & \multicolumn{1}{l|}{0.392957} & \multicolumn{1}{l|}{0.395358} & \multicolumn{1}{l|}{0.402319} & \multicolumn{1}{l|}{0.413203} & \multicolumn{1}{l|}{0.443385} & \multicolumn{1}{l|}{0.538069} & 0.721063 \\ \hline
$Q=0.25$ & \multicolumn{1}{l|}{0.369174} & \multicolumn{1}{l|}{0.370338} & \multicolumn{1}{l|}{0.373793} & \multicolumn{1}{l|}{0.379439} & \multicolumn{1}{l|}{0.396583} & \multicolumn{1}{l|}{0.461671} & 0.611542 \\ \hline
$Q=0.30$ & \multicolumn{1}{l|}{0.322735} & \multicolumn{1}{l|}{0.323114} & \multicolumn{1}{l|}{0.324264} & \multicolumn{1}{l|}{0.326222} & \multicolumn{1}{l|}{0.332814} & \multicolumn{1}{l|}{0.36824} & 0.493662 \\ \hline
\end{tabular}
\caption{{Variations of the orbital equilibrium position for different charge values and different angular momenta.}}
\label{tab:aos1}
\end{table}

\begin{table}[]
\begin{tabular}{|l|l|}
\hline
       &$r_{0}$       \\ \hline
Q=0.15 & 0.407356 \\ \hline
Q=0.25 & 0.392583 \\ \hline
Q=0.25 & 0.369016 \\ \hline
Q=0.3  & 0.322696 \\ \hline
\end{tabular}
\caption{{The black hole horizon at different charge values.}}
\label{tab:aos2}
\end{table}

\begin{figure}
\begin{center}
\subfigure[{$Q=0.15$.}]{
\includegraphics[width=0.45\textwidth]{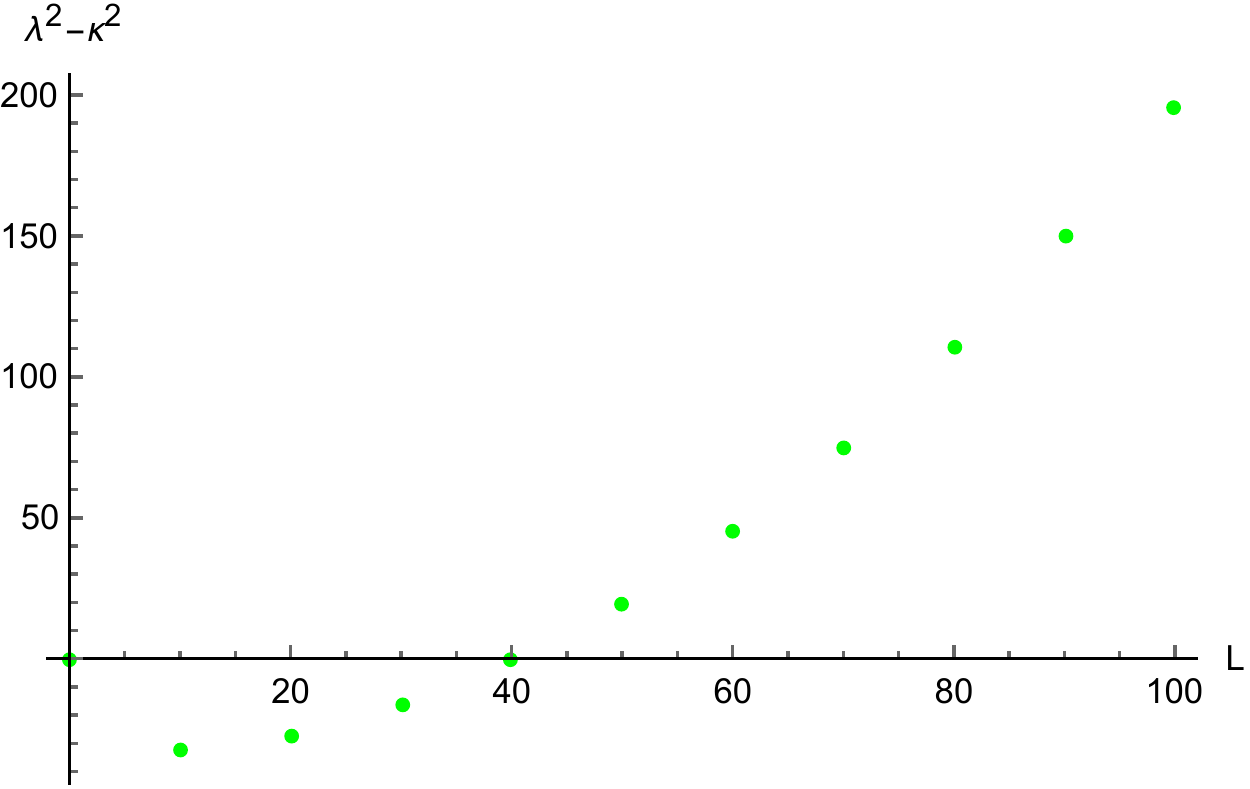}\label{fig:01}}
\subfigure[{$Q=0.20$.}]{
\includegraphics[width=0.45\textwidth]{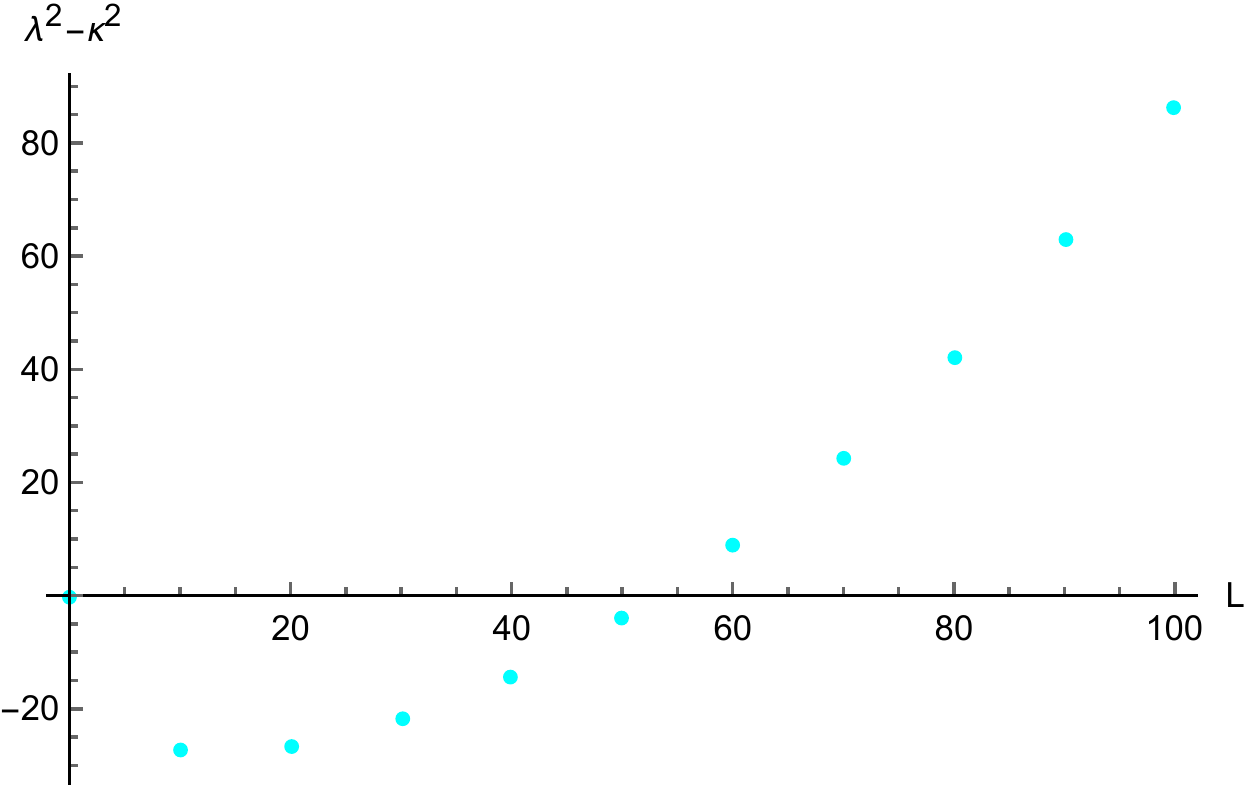}\label{fig:02}}
\subfigure[{$Q=0.25$.}]{
\includegraphics[width=0.45\textwidth]{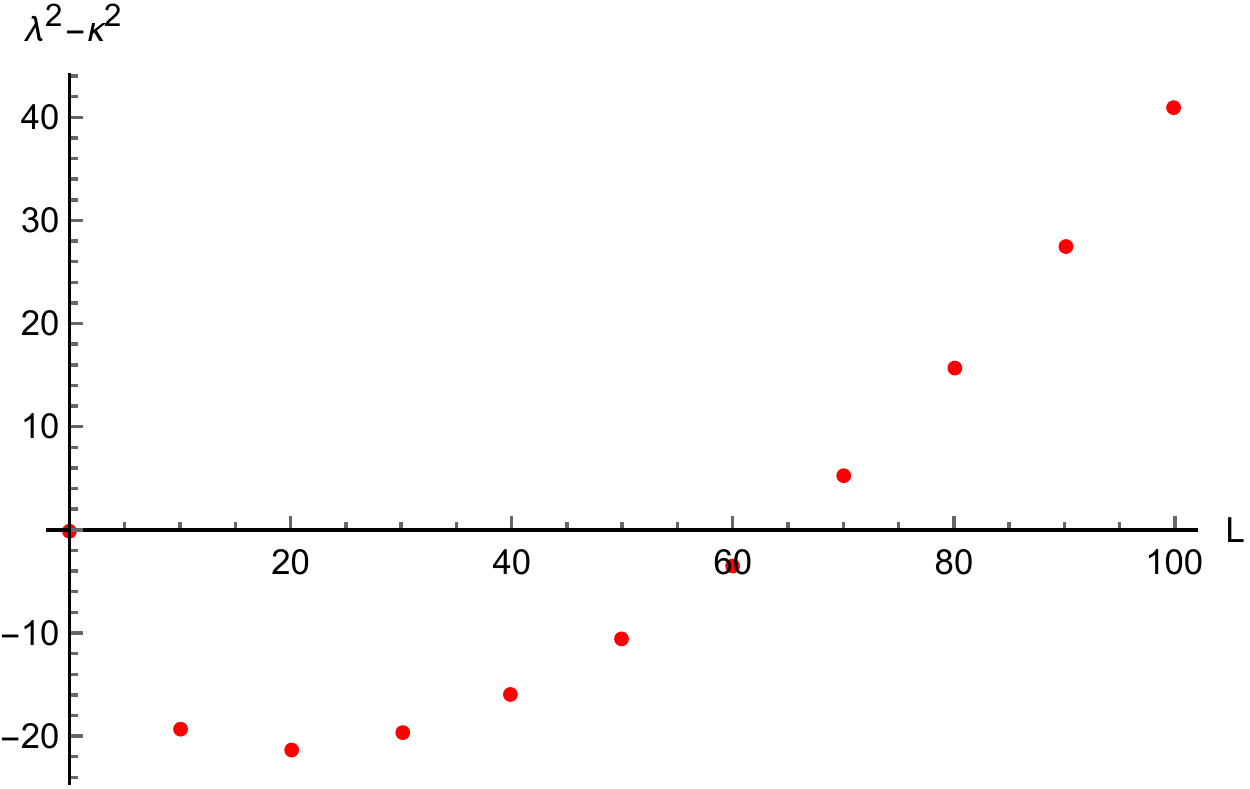}\label{fig:03}}
\subfigure[{$Q=0.30$.}]{
\includegraphics[width=0.45\textwidth]{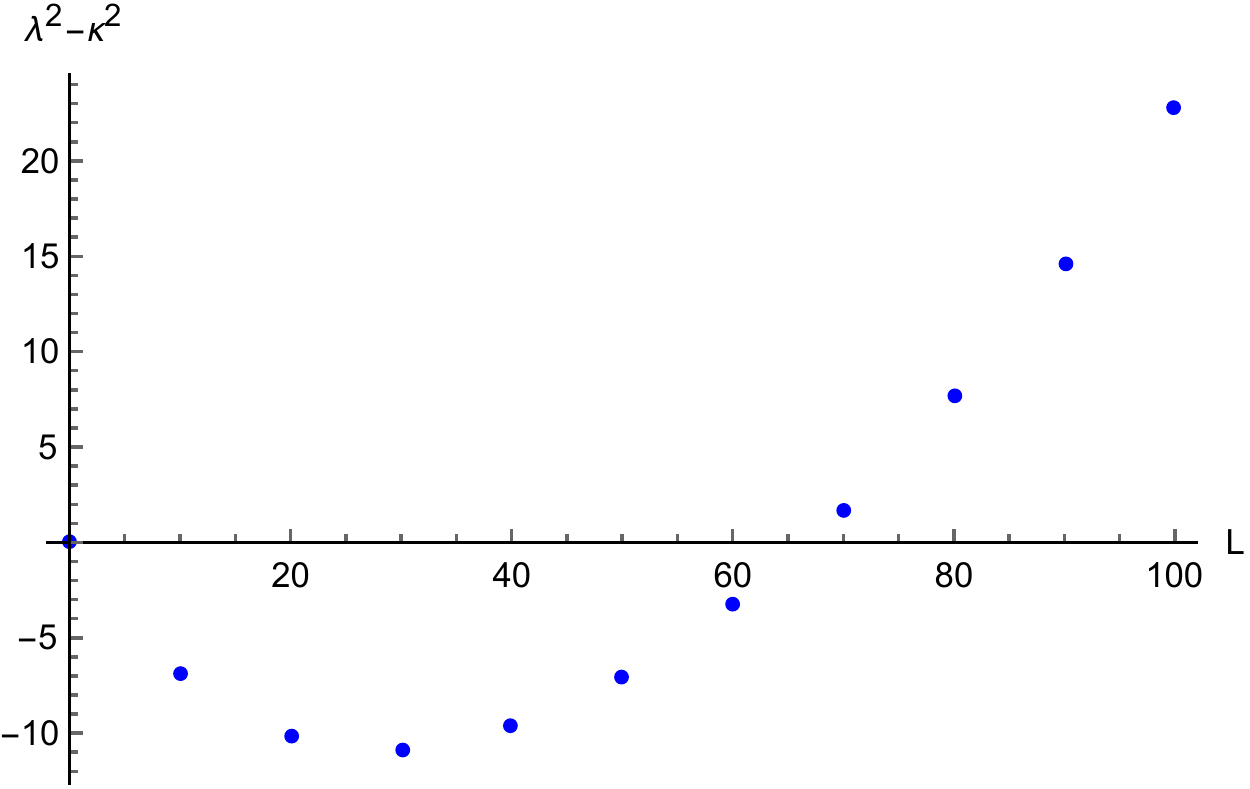}\label{fig:04}}
\end{center}
\caption{The influence of the angular momentum on the chaos bound.}%
\label{fig:J00}
\end{figure}

The metric of the torus-like black hole in four dimensional space is given as follows \cite{intro-Huang:1995zb, intro-Han:2019kjr}
\begin{equation}
F(r)=-\frac{2M}{\pi r}+\frac{4Q^{2}}{\pi r^{2}}-\frac{\varLambda r^{2}}{3},
\label{eqn:P31}
\end{equation}
\begin{equation}
C(r)=r^{2},
\label{eqn:P32}
\end{equation}
\begin{equation}
D(r)=r^{2}\sin^{2}\theta,
\label{eqn:P33}
\end{equation}
where, $M$ and $Q$ are the mass and electric charge of the black hole, respectively. $\varLambda$ is the cosmological constant, which is connected with a thermodynamic variable ($P=-\frac{\Lambda}{8\pi}=\frac{3}{8\text{\ensuremath{\pi}}l^{2}}$). In case of $\varLambda<0$, the coordinate singularities appear at the horizon radii ($r_{\pm}$), meanwhile $F(r_{\pm})=0$. The nonvanishing component of electromagnetic vector potential for the torus-like black hole in four dimensional space is \cite{intro-Han:2019kjr}
\begin{equation}
A_{t}=-\frac{4Q}{r}.
\label{eqn:P34}
\end{equation}

According to the literature \cite{Gao:2022ybw}, here we set the electromagnetic vector potential $A_{t}$ to $\frac{4Q}{r}$.

With the help of the above, we then focus on finding the positions of the equilibrium orbits. In the equatorial plane, $\theta$ is set to $\frac{\pi}{2}$.
In the case $\pi_{r}=\frac{d\pi_{r}}{dr}=0$, the expression for the equilibrium position of the orbit can be yielded. And the expression of $\frac{d\pi_{r}}{dt}$ is
\begin{equation}\label{eqn:P35}
  \frac{d\pi_{r}}{dt}=\frac{4qQ}{r^{2}}+\frac{3L^{2}(8Q^{2}-3Mr)+r^{2}(12Q^{2}-3Mr-8P\pi^{2}r^{4})}{\sqrt{6\pi}r^{3}\sqrt{(L^{2}+r^{2})(6Q^{2}-3Mr+4P\pi^{2}r^{4})}}.
\end{equation}

During the numerical analysis, $P=1,M=1,q=15$ is set, and the change of the equilibrium orbit position is observed by changing the value of $Q$ and $L$. The specific numerical changes are summarized in the Table \ref{tab:aos1}.

It can be clearly observed from the table that when the charge $Q$ is fixed, the value of $r_{0}$ increases gradually with the increase of angular momentum. In order to compare the orbital equilibrium position with the position of the black hole horizon, Table \ref{tab:aos2} gives the position of the black hole horizon for different charges. By observation, it can be intuitively found that the positions of the orbits gradually deviate from the black hole horizon. When only the charge is considered, it is observed that as the charge increases, the value of the equilibrium position of the orbit gradually decreases.

Further exploration of the variation of the black hole chaos bound is in the near-horizon region and at a certain distance from the horizon. For the observation, we perform numerical calculations for the Lyapunov exponent at the equilibrium orbits and the surface gravity at the black hole horizon. The results of numerical calculations are used to plot Fig. \ref{fig:J00}.

As can be observed from Fig. \ref{fig:J00}, there are both violations of the bounds. When the charge value is fixed, the variation of the angular momentum has a significant effect on the value of the Lyapunov exponent, which leads to the change of the bound. There is no violation when the angular momentum is relatively small, i.e., the position of the equilibrium orbit is close to the black hole horizon. As the angular momentum increases, the bound is violated at a certain distance from the black hole horizon. There exists a special value of angular momentum, when the angular momentum is larger than this value, the violation phenomenon appears. It is observed that this special value increases with the increase of the charge value.

\section{Conclusions}
\label{sec:D}
In this paper, the variations of the chaos bound of the torus-like black hole in the near-horizon region and in the region at a certain distance from the black hole horizon were studied. The effects of different charges and different angular momenta on the equilibrium position of the orbit were first investigated. It is found that when the charge value is fixed, the orbital equilibrium position gradually moves away from the black hole horizon with the increase of angular momentum. When only the charge parameter is considered, the orbital equilibrium position parameter decreases with the increase of charge value, which is the same as the effect of charge on the black hole horizon.

Subsequently, the Liapunov exponent is calculated by the Jacobi matrix, and further study reveals that the angular momentum has a great influence on the value of Liapunov exponent. The plot shows that the chaos bound is not violated in the near-horizon region; and with the increase of angular momentum, the bound is violated at a certain distance from the black hole horizon. The conclusion is the same as the one reached in Ref. \cite{Gao:2022ybw}.

\begin{acknowledgments}
We are grateful to  Deyou Chen, Peng Wang, Haitang Yang, Jun Tao and Xiaobo Guo for useful discussions. The authors contributed equally to this work. This work is supported in part by NSFC (Grant No. 11747171), Xinglin Scholars Project of Chengdu University of Traditional Chinese Medicine (Grant no.QNXZ2018050).
\end{acknowledgments}

\end{document}